\documentstyle[twoside,fleqn,espcrc2]{article}

\newcommand{\AmS}{{\protect\the\textfont2
  A\kern-.1667em\lower.5ex\hbox{M}\kern-.125emS}}

\title{The Phase Diagrams of the Schwinger  and Gross-Neveu
       Models with Wilson Fermions}

\author{R. Kenna\address{School of Mathematics, Trinity College Dublin, 
        Ireland,}%
        \thanks{Supported by EU TMR
           Project No. ERBFMBI-CT96-1757 and a Forbairt European Presidency 
           Post Doctoral Fellowship.}
        C. Pinto${}^{\rm a,}{}$\address{
        Hitachi Dublin Laboratory, Dublin, Ireland}
        and J.C. Sexton${}^{\rm a,b}{}$}
       
\begin{document}

\begin{abstract}
A new method to analytically determine the partition function zeroes
of weakly coupled theories on finite--size lattices is developed.
Applied to the lattice Schwinger model, this reveals the possible
absence of a phase transition at fixed weak coupling.
We show how finite--size scaling techniques on small
or moderate lattice sizes may mimic the
presence of a spurious phase transition.
Application of our method to the Gross--Neveu model yields
a phase diagram consistent with that coming from a saddle point analysis.
\end{abstract}

\maketitle

\section{Introduction}

There has recently been considerable discussion on the phase structure 
of theories with Wilson fermions. A system of free Wilson fermions 
exhibits a second order phase transition at $\kappa=1/2d$, $\kappa$ being
the hopping parameter and $d$  the dimensionality. 
Discussions concern the extent to which this phase transition persists 
when a bosonic (e.g. gauge) field is included. In the Schwinger
model, with inverse gauge coupling squared $\beta$ , the expectation 
is that there is a line of phase transitions in the $(\beta,\kappa)$ 
plane extending to the strong coupling limit 
$\beta = 0$ \cite{strong}. This critical line is also 
expected to recover massless physics and therein lies the importance 
in determining its nature and position \cite{GaLa94,AzDi96}

The phase diagram of the Schwinger model has been numerically
determined by two groups, which, while in rough agreement regarding the 
location of the critical line, differ in their conclusions regarding 
its quantitive critical properties. Using Lee--Yang zeroes,
finite--size scaling and other techniques the results of 
\cite{GaLa94} support the free boson scenario, where the
model lies in the same universality class as the Ising model,
with critical  exponents $\nu = 1$, $\alpha = 0$. This is not in
agreement with \cite{AzDi96} where finite--size scaling  of Lee--Yang 
zeroes on larger lattices provides evidence for $\nu = 2/3$, 
$\alpha = 2/3$.

We analytically determine
the phase structure in the weakly coupled regime. We find that the 
zeroes in the Schwinger model with fixed weak gauge coupling
display unusual  behaviour, do not accumulate on 
the real axis and may not, in fact, lead to a phase 
transition. The behaviour of these zeroes is, however, such as to mimic 
the appearance of a phase transition when a finite--size analysis is 
restricted to small or moderate lattices.

Applying our method to the $d=2$ Gross--Neveu model yields results
consistent with Aoki's saddle point analysis \cite{Aoki} and this consistency
adds confidence to our new analytic approach.

\section{Lattice Schwinger Model}

We consider a  $d=2$ lattice of spacing $a$ and $N$ 
sites in each direction. For the fermion fields, we impose antiperiodic 
(periodic) boundary conditions in the $1$- ($2$-) direction.
The Fourier transformed fermion
fields then have momenta $p_\mu = (2 \pi /N a) \hat{p}_\mu$ where $\hat{p}_1$
is half--integer and $\hat{p}_2$ is integer. The action is
$ S_{\rm{QED}} = S_G + S_F^{(W)}$,
where $
 S_F^{(W)} = \sum_{m,n} 
\bar{\psi}_m M^{(W)}_{m,n}  \psi_n
 $ and
\begin{eqnarray}
\lefteqn{
          M^{(W)}_{m,n} =
          \frac{\delta_{m,n}}{2\kappa}
          -
          \frac{1}{2}\sum_\mu{
                               \left\{
                                      ( 1 - \gamma_\mu ) U_\mu(m)
                                      \delta_{m+\hat{\mu},n}
                               \right.
                             }
        }
\nonumber
\\
& &
        {
         \left.
                + ( 1 + \gamma_\mu ) 
                U^\dagger_\mu(n)
                \delta_{m-\hat{\mu},n}
          \right\}
}
\quad .
\label{3.5}
\end{eqnarray}
This  fermion matrix
consists of free and interacting parts,
$M^{(W)} = M^{(0)} + M^{({\rm{int}})} $ with
$M^{(0)}$  given by $U_\mu = 1$ in (\ref{3.5}).
Here
$U_\mu (n) = \exp{(i e_0 a A_\mu (n))}$
is the link variable. 
The partition function is $
 Z = 
 \int{
      {\cal{D}} U
      \exp{(
         \{
           -S_{\rm{G}} -S^{(W)}_{\rm{F}}
         \}
        )}
     }
 \langle
       \det{M^{(W)}}
 \rangle
$,
the integration over  fermions having been 
performed.
The gauge action is
$
 S_G=
    \beta\sum_P{\left[1-\frac{1}{2}(U_P+U_P^\dagger)\right]}
$,
where $U_P$ is the  product of link variables around a 
plaquette, $\beta = 1/e_0^2 a^2$
and $\langle {\cal{O}} \rangle $ is a pure gauge expectation.

For $\beta = \infty$, the partition function is
simply proportional to $ \det{M^{(0)}} $. 
Then the free partition function can
be written in terms of its  eigenvalues
$
 \lambda^{(0)}_\alpha ({\hat{p}}) 
 = 
 1/2\kappa
 - \sum_{\mu=1}^2  \cos{p_\mu a} 
 + i (-)^\alpha  \sqrt{ \sum_{\mu=1}^2{\sin^2{p_\mu a}}}
$ 
which are either 2--fold ($\hat{p}_2 = 0$ or $-N/2$) or 4--fold 
degenerate.
In this case, and with $\eta = 1/2\kappa$, the zeroes are 
 $\eta^{(0)}_\alpha(\hat{p}) =
 \sum_{\mu=1}^2 r \cos{p_\mu a} 
 - i (-)^\alpha  \sqrt{\sum_{\mu=1}^2{\sin{p_\mu a}}}
$.
The lowest zero with finite real part in $\kappa$ has
$ \hat{p}= (\pm 1/2,0)$.
Pinching of the positive real finite
hopping parameter axis occurs at 
$\kappa_c = 1/2d $ and application of finite--size scaling to
the imaginary parts of the zero 
gives the critical exponent $ \nu = 1$ ($\alpha = 0$ then follows from
hyperscaling). Therefore the free fermion model is in the same
universality class as the Ising model in two dimensions and
describes free bosons.

\section{The Weak Coupling Expansion}

The weak coupling expansion is obtained by expanding the link
variables $U_\mu (n)$ as a power series in $e_0$.
The fermion determinant 
can then be expanded giving the
`traditional' form for the weak coupling expansion
(see e.g., \cite{Rothe}).
This expansion is analytic in $\eta$ with
poles at $\eta = \eta_i^{(0)}$ \cite{us}. 

An alternative formulation of the partition function may be obtained
by writing the  fermion matrix as $M^{(W)} = \eta 
+ H $. The 
determinant $\det M^{(W)} = \det(\eta + H)$ is a polynomial
in $\eta$ for finite lattice size.
 Therefore its pure gauge expectation value
is also a polynomial in $\eta$ and
may be written
$
 \langle \det M^{(W)} \rangle = \prod_{i=1}^{dN^d}
 (\eta - \eta_i)
$. Here $\eta_i$ represent $\eta_\alpha ({\hat{p}})$ and are the
zeroes
in the complex $1/2 \kappa$ plane.
We may write a `multiplicative' weak coupling expansion as
\begin{equation}
 \frac{ 
     \langle  \det{M^{(W)}} \rangle 
      }{
       \det{M^{(0)}}
      }
 = \prod_{i=1}^{dN^d}\left(
  1 - \frac{\Delta_i}{\eta - \eta_i^{(0)}}
                    \right)
\quad ,
\label{multiplicative}
\end{equation}
where $\Delta_i = \eta_i - \eta_i^{(0)}$ are the shifts that occur
in the zeroes when the gauge field is turned on.
Note  that (\ref{multiplicative}) is analytic in $\eta$ with
poles at $\eta_i^{(0)}$.
Since the multiplicative expression (\ref{multiplicative}) 
must be the same as the `traditional' weak coupling expansion,
the residues of the poles
must be equal.

Equating these residues gives the the shifts in the positions 
of the zeroes in the two fold degenerate case
(the case of four fold degeneracies is more complicated but we expect 
the lowest zeroes to be erstwhile two fold degenerate).
In this way, the shifts in the positions of the lowest  zeroes are 
determined to ${\cal{O}}(e^2)$. Our calculation of pure gauge expectation 
values is done in Feynman gauge.

\section{Results and Conclusions}

\begin{figure}[htb]
\vspace{3.5cm}
\includegraphics{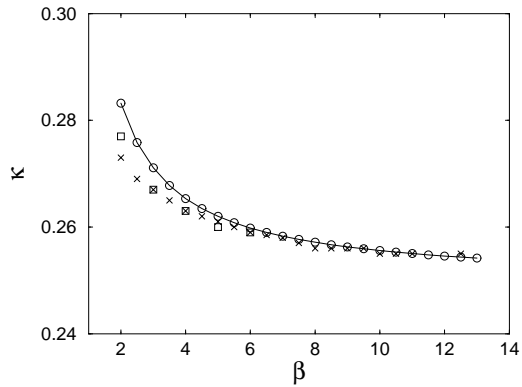}
\caption{The would-be phase diagram coming from the real parts of the 
lowest zeroes at $N=24$.}
\label{fig:1}
\end{figure}
\vspace{-0.5cm}
A standard numerical technique for
determination of the phase diagram 
is to approximate the  critical point by 
the real part of the lowest zero  for some
large lattice. Plotted against $\beta$ this
approximates the phase diagram.
In Figure~1 we present such a plot for $N=24$ (circles)
to compare with the results of \cite{AzDi96} (crosses) also at $N=24$.
The phase diagram of \cite{GaLa94}
for the Schwinger model with {\em{two}} fermion flavours coming from
a separate PCAC based analysis is also included
(squares)
for comparison.

Figure~2 is a finite--size scaling plot for the lowest zero 
at $\beta=10$ and $N=8$ -- $62$ coming from our weak coupling 
analysis (circles). The corresponding data from the numerical analysis 
of \cite{AzDi96} are also included (crosses). The two  lines are
linear fits to the first and second zeroes for lattice sizes $16,20,24$ 
which are those analysed in  \cite{AzDi96}. These yield slopes $-1.5$ 
and $-1.4$ respectively, corresponding to $\nu \approx 2/3$.
A similar plot may be made to compare with the numerical results of
\cite{GaLa94} who use $\beta = 5$  and $N=2,4,8$. 
There, the best linear fit gives $\nu$ compatable with $1$.
Fitting to our small lattice data also gives $\nu \approx 1$.
\begin{figure}[htb]
\vspace{4.0cm}
\includegraphics{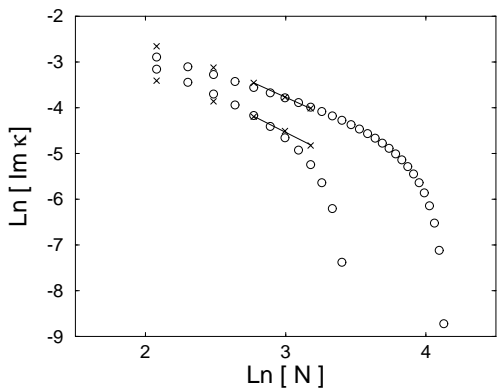}
\caption{Finite size scaling of the imaginary parts of the first
two Lee--Yang zeroes at $\beta=10$.}
\label{fig:2}
\end{figure}
\vspace{-0.5cm}
Thus we find that confining the finite--size scaling
analysis to the range $N=2$ -- $8$ yields
 $\nu=1$ in agreement with \cite{GaLa94}
while a  corresponding analysis with
$N=8$--$24$ gives $\nu=2/3$
in agreement with \cite{AzDi96}.
It is clear from the figure, however, that
the curve does not in fact settle to a  finite--size scaling line.
Instead as $N$ increases, the lowest zeroes cross the
real axis. The first two zeroes therefore fail to accumulate 
and do not contribute to critical behaviour \cite{Abe}.
These zeroes 
are isolated singularities,
having measure zero in the thermodynamic limit. 
Although the possibility of existence of isolated singularities
and non--accumulation of partition function zeroes has been
known for a long time \cite{Abe}, this is to our knowledge 
the first instance 
where such behaviour has been observed. 

We have also applied our new analytic technique to the two dimensional
Gross--Neveu model \cite{us2}. There we find the expected accumulation of
zeroes and a weakly coupled phase diagram  which is consistent
with that determined by Aoki using a saddle point approach \cite{Aoki}.

In conclusion, we have developed a new method to analytically
determine the Lee--Yang zeroes of weakly coupled theories.
In the free case, there is a phase transition
precipitated by the accumulation of Lee--Yang zeroes on the 
real hopping parameter axis. 
In the weakly coupled lattice Schwinger model at fixed $\beta$, 
this accumulation no longer occurs for the first couple of zeroes. 
Instead, the movement of these zeroes for  small and
moderately sized lattices mimics phase transition like behaviour.
As the lattice size becomes large, however, 
these zeroes move across the real axis, 
and do not give rise to a phase transition.
In the Gross--Neveu case, our phase diagram is consistent with Aoki's
saddle point analysis \cite{us2}.


\begin{thebibliography}{9}
\bibitem{strong}
M. Salmhofer, Nucl. Phys. B 362 (1991) 641;
H. Gausterer, C.B. Lang and M. Salmhofer, Nucl. Phys. B
388 (1992) 275;
H. Gausterer and C.B. Lang, Nucl. Phys. B 455 1995 785;
F. Karsch, E. Meggiolaro and  L. Turko
Phys. Rev. D 51 (1995) 6417;
C. Gattringer, hep-lat/9903021 (to appear in Nucl. Phys. B).
\bibitem{GaLa94}
H. Gausterer and C.B. Lang, Phys. Lett. B 341 (1994) 46;
Nucl. Phys. B (Proc. Supl.) 34 (1994) 201;
I. Hip, C.B. Lang and R. Teppner,  
Nucl. Phys. B (Proc. Supl.) 63 (1998) 682.
\bibitem{AzDi96}
V. Azcoiti, G. Di Carlo, A. Galante, A.F. Grillo and V. Laliena,
Phys. Rev. D 50 (1994) 6994; ibid. 53 (1996) 5069.
\bibitem{Aoki}
S. Aoki, Phys. Rev. D 30 (1984) 2653.
\bibitem{Rothe}
H.J. Rothe, {\em{Lattice Gauge Theories}} (World Scientific, Singapore,
1997).
\bibitem{us}
 R. Kenna, J.C. Sexton and C. Pinto, hep-lat/9812004.
\bibitem{Abe}
R. Abe, Prog. Theor. Phys. 38 (1967) 322;
M. Salmhofer, Nucl. Phys. B, Proc. Suppl. 30 (1993) 81;
Helv. Phys. Acta 67 (1994) 257.
\bibitem{us2}
 R. Kenna, J.C. Sexton and C. Pinto, in preparation.
\end{thebibliography}
\end{document}